\documentclass[preprint]{elsarticle}

\usepackage{lineno,hyperref}
\modulolinenumbers[10]

\usepackage{color}

\begin{document}

\begin{frontmatter}

\title{Machine learning the dynamics of quantum kicked rotor}

\author{ Tomohiro Mano}
\author{Tomi Ohtsuki}
\address{Physics Division, Sophia University, Kioicho 7-1, Chiyoda-ku, Tokyo 102-8554, Japan}
\ead{ohtsuki@sophia.ac.jp}




\begin{abstract}
Using the multilayer convolutional neural network (CNN), we can detect the quantum phases in random electron systems, and phase diagrams of two and higher dimensional Anderson transitions and quantum percolations as well as disordered topological systems have been obtained.  Here, instead of using CNN to analyze the wave functions,
we analyze the dynamics of wave packets via long short-term memory network (LSTM).
We adopt the quasi-periodic quantum kicked rotors,
which simulate the three and four dimensional Anderson transitions.
By supervised training, we let LSTM extract the features of the time series of wave packet displacements
in localized and delocalized phases.
 We then simulate the wave packets in unknown phases and let LSTM classify
the time series to localized and delocalized phases.
We compare the phase diagrams obtained by LSTM and those obtained by CNN.
\end{abstract}


\end{frontmatter}

\linenumbers

\section{Introduction}
Critical behaviors of the Anderson transition\cite{Anderson58} have been attracting considerable
attention for more than half a century.
The problem is related to quantum percolation\cite{Kirkpatrick72,Sur76,Schubert05,StaufferBook,Ujfalusi15},
where the wave functions on the
randomly connected lattice sites begin to be extended \cite{Makiuchi18}.
Electron states on random lattice systems are difficult to study, because
 the conventional methods of using the transfer
matrix\cite{Slevin14} are not applicable.
The scaling analyses of the energy level statistics\cite{Shklovskii93} are also difficult, if not impossible\cite{Berkovits96,Kaneko99},
owing to the spiky density of states \cite{Ujfalusi15}.

To overcome these difficulties, neural networks\cite{Mehta19,Carleo19,Ohtsuki20,Bedolla20} to classify
the states \cite{Tomoki16,Tomoki17,Broecker16,Carrasquilla17,Zhang16,Zhang17b,Yoshioka18,Nieuwenburg17,Zhang18,Araki19,Carvalho18} had been used.
That is, instead of classifying images of photos, we input the wave functions (actually the squared modulus
of them) at the Fermi energy
and classify them to metal, insulator, topological insulator etc.
First we train the convolutional neural network in Anderson model of localization, whose
phase diagram is well known.  We then  apply the CNN to classify the eigenfunctions
of quantum percolation to metal or insulator relying on the generalization capability of CNN.
 We have shown in refs.~\cite{Mano17,Mano19} that this method
is free from the above difficulties and works well in determining the phase diagrams of quantum percolation.

The above method, however, requires many eigenfunctions, which are difficult to obtain in higher dimensions.
One of the way to study the Anderson transition without diagonalizing the Hamiltonian
 is to use quantum kicked rotor (QKR) \cite{Chabe08,Lemarie10,Lopez12},
where we analyze the wave packet dynamics in one dimension.
The simple quantum kicked rotor can be mapped to one-dimensional (1D) Anderson model\cite{haakebook},
whereas
with incommensurate modulation of the strength of kick, the model is mapped
to tight binding models in higher dimensions\cite{Casati89}.

In this paper, we draw phase diagrams of the three dimensional (3D) and four dimensional (4D)
tight binding models that correspond to the QKR using the CNN
trained for standard Anderson models of localization.
We also analyze the time series of QKR via long short-term memory (LSTM) network\cite{Hochreiter97,Ola15},
let LSTM classify the time series to localized/delocalized phases,
draw the phase diagrams, and compare them with
those drawn by the CNN analyses of tight binding models.
We demonstrate that the phase boundaries of localized and delocalized phases
are less noisy in the case of LSTM.

\section{Model and Method}
We consider QKR with incommensurate modulation of the kick as follows;
\begin{equation}
H(t)=\frac{p^2}{2}+K\cos x \times \sum_n \delta(t-n)\times  F(t)\,,
\end{equation}
with 
\begin{equation}
\label{eq:modulation}
F(t)=\left\{
\begin{array}{ll}
1+\epsilon \cos(\omega_2 t+\theta_2)\times \cos(\omega_3 t+\theta_3) & \mathrm{3D}\\
1+\epsilon \cos(\omega_2 t+\theta_2)\times \cos(\omega_3 t+\theta_3)\times \cos(\omega_4 t+\theta_4)& \mathrm{4D}
\end{array}
\right. \,
\end{equation}
where $\omega_i$ are irrational numbers that are incommensurate with each other,
$K$ the strength of the kick, $\epsilon$ the modulation strength, and $\theta_i$ the initial phases.
We took $\omega_2=2\pi \sqrt{5}, \omega_3=2\pi \sqrt{13}$ and $\omega_4=2\pi \sqrt{23}$
\cite{Chabe08,Lemarie10,Lopez12}.

We analyze the QKR in two ways. One way is to map this model to tight binding models,
\begin{equation}
\label{eq:tb}
H_\mathrm{tb}=\sum_m \varepsilon_m |m\rangle \langle m|+\sum_{m,r}W_r|m\rangle \langle m-r| \,,
\end{equation}
with $m=(m_1, m_2, m_3)$, %
$\epsilon_m =\tan\left[-\frac{1}{2}(m_1^2\hbar/2+\omega_2 m_2 +\omega_3 m_3)\right]$
 and 
 $W_{r_1,r_2,r_3}$ the Fourier transform of 
 $W(x_1,x_2,x_3)=\tan\left[\frac{K \cos x_1(1+\epsilon\cos x_2\cos x_3)}{2\hbar} \right]$
 for 3D \cite{Lemarie10,haakebook}.  For 4D, we include $m_4, r_4$ and $x_4$ in a
 straightforward way.
We diagonalize $H_\mathrm{tb}$ numerically to obtain the eigenfunctions, and
let CNN determine whether they are localized or delocalized.
Details of the CNN method is reviewed in ref.~\cite{Ohtsuki20}.
Note that $H_\mathrm{tb}$ is defined on 3D cubic lattice for the case of
two incommensurate frequencies ($\omega_2$ and $\omega_3$), whereas it is on 4D hypercubic
lattice in the case of three incommensurate frequencies ($\omega_2, \omega_3,$ and $\omega_4$).
Note also that we use CNN that has been trained for Anderson models of localization \cite{Mano17,Ohtsuki20}.

The other way is to solve the time dependent Schr\"odinger equation
\begin{equation}
i\hbar \frac{d}{dt}\psi (t)=H(t)\psi(t)\,,
\end{equation}
with $\hbar$ set to 2.89\cite{Chabe08,Lemarie10,Lopez12},
calculate the time dependence of ``displacement" in momentum space,
\begin{equation}
p^2(t)=\langle\psi(t) |p^2|\psi(t)\rangle\,,
\end{equation}
and analyze the time series of $p^2(t)$ via LSTM.
To follow the wave packet time evolution from $t=n+\eta$ to $t=n+1+\eta$ with
$\eta$ infinitely small positive number,
we use
\begin{equation}
\psi(n+1)=\exp\left(-i\frac{K\, F(n+1)\, \cos x}{\hbar}\right) \times \exp\left(-i\frac{p^2}{2\hbar}\right)\psi(n)\,.
\end{equation}
We work in the $p$-space, and the multiplication of $\exp(-i a \cos x)$ ($a=K F(n+1)/\hbar$)
is expanded in the
$p$-space as $\langle p |\exp(-i a \cos x)|p'\rangle$, which is expressed by Bessel functions.

\begin{figure}[htbp]
  \begin{center}
   \includegraphics[angle=0,width=1.\textwidth]{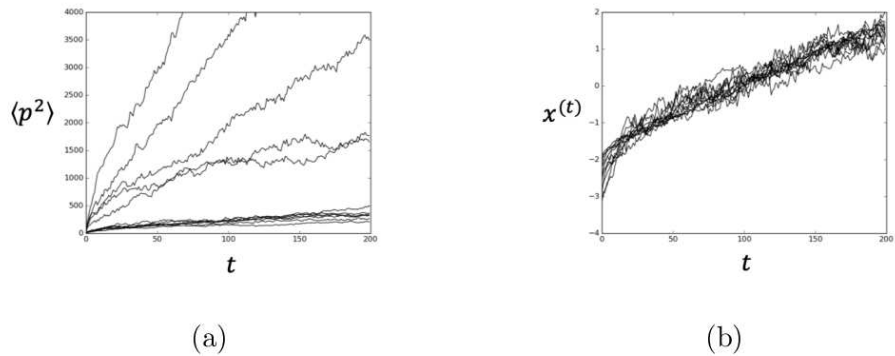}
  \end{center}
  \vspace{-1.5cm}
  \caption{$p^2(t)$ vs. $t$ for various $K$ and $\epsilon$ with two
  incommensurate frequencies (3D case).
  (a) the plot before normalization.
  $p^2(t)$ is proportional to $t$ for delocalized states, whereas
  it saturates to finite values for localized states.
  (b) after normalizing $p^2(t)$ to $x^{(t)}$, the mean and standard deviation of
  which are 0 and 1, respectively.
  We have calculated the wave packet dynamics up to $T=10^4$ time steps,
  and recorded $p^2(t)$ at every 50 time steps.
  }
\label{fig:p2t}
\end{figure}

\section{Results}
\begin{figure}[htbp]
  \begin{center}
     \begin{tabular}{cc}    
 \begin{minipage}{0.5\hsize}
  \begin{center}
   \includegraphics[angle=0,width=\textwidth]{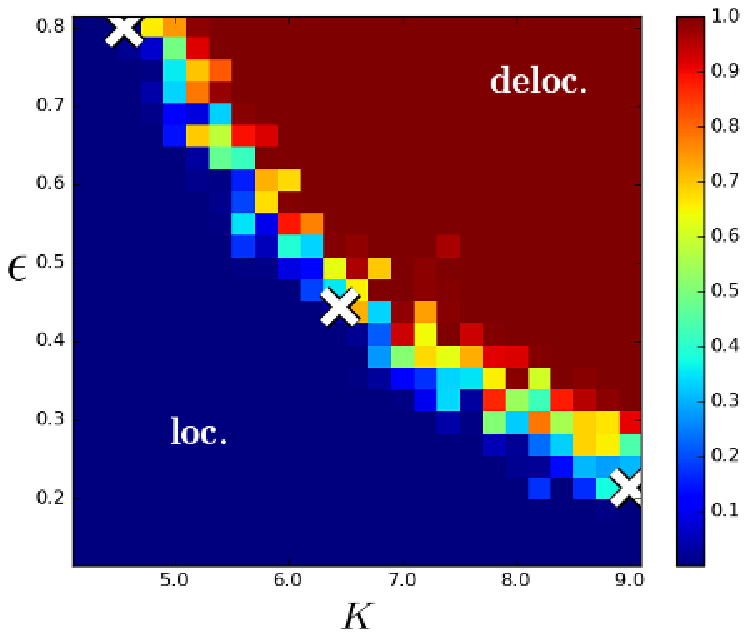}
      \hspace{1.6cm} (a) 3D CNN
     \end{center}
 \end{minipage}
 \begin{minipage}{0.5\hsize}
  \begin{center}
  \includegraphics[angle=0,width=\textwidth]{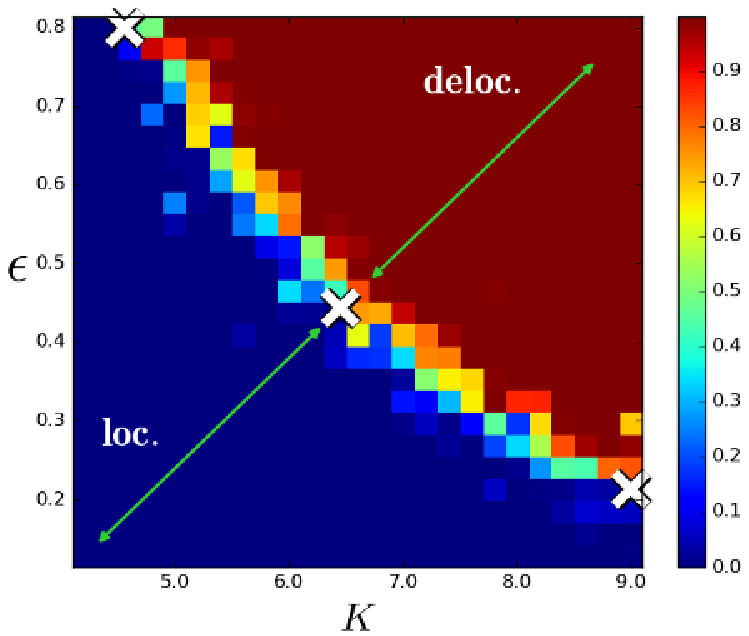}
      \hspace{1.6cm} (b) 3D LSTM
     \end{center}
 \end{minipage}\\
 
 \begin{minipage}{0.5\hsize}
  \begin{center}
  \includegraphics[angle=0,width=\textwidth]{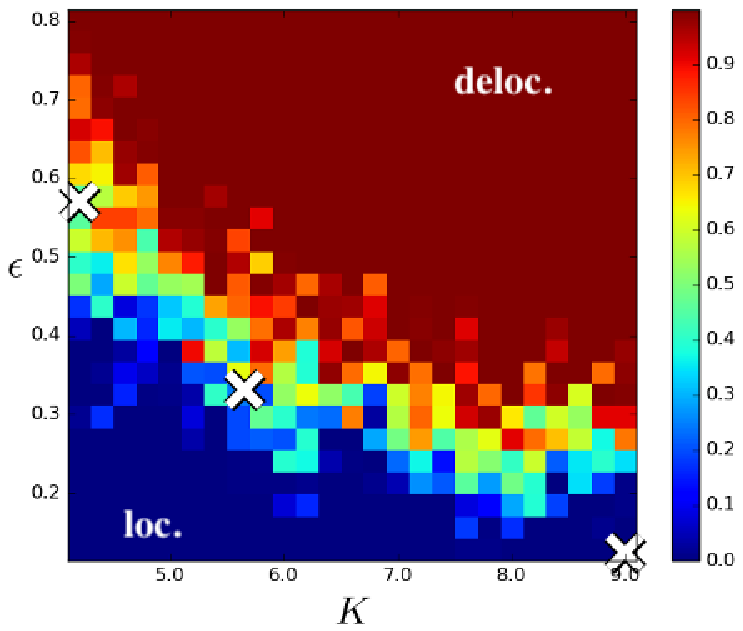}
      \hspace{1.6cm} (c) 4D CNN
     \end{center}
 \end{minipage}
 \begin{minipage}{0.5\hsize}
  \begin{center}
  \includegraphics[angle=0,width=\textwidth]{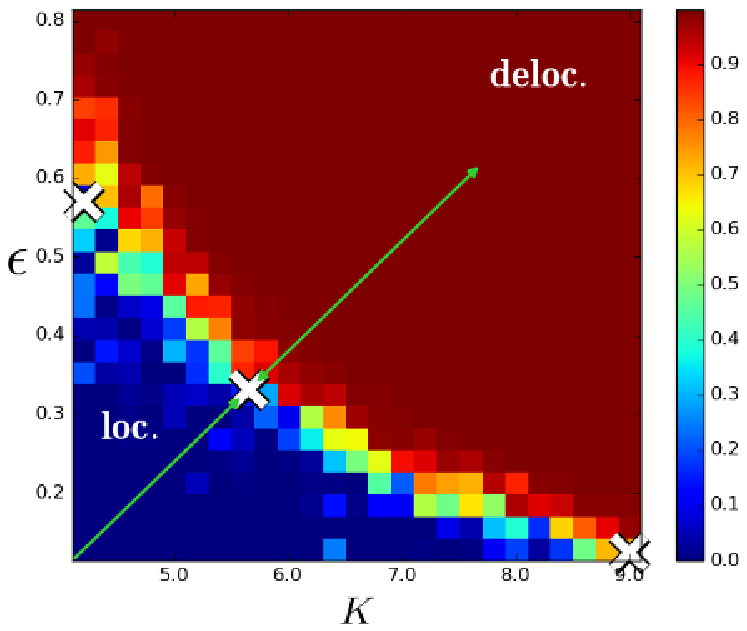}
      \hspace{1.6cm} (d) 4D LSTM
     \end{center}
 \end{minipage}
  \end{tabular}
  \end{center}
  \caption{Phase diagrams of QKR in $\epsilon$-$K$ plane.
 At each $(K,\epsilon)$, we plot the probability that
 the parameter belongs to the delocalized phase.
 Those obtained by CNN[(a) and (c)]
and those by LSTM[(b) and (d)].
3D cases [(a) and (b)] and 4D cases[(c) and (d)].  
The CNN is trained by Anderson model of localization.
The training regions of LSTM are indicated as green arrows.
White crosses are obtained by the critical behaviors of $p^2(t)$.
In all cases, average over 5 samples has been taken.
Random choice of $0\le\theta_i<2\pi$ and random shift of 
$(m_1, m_2,\cdots)\rightarrow (m_1+\beta_1,m_2+\beta_2,\cdots)\,,\,0\le\beta_i<1$
have been performed\cite{Lemarie10}. 
  }
  \label{fig:phaseDiagrams}
\end{figure}

We first apply the CNN trained for the Anderson model to
the wave functions obtained by diagonalizing Eq.~(\ref{eq:tb}).
For 3D systems, the system size is $32\times 32\times 32$, whereas for 4D
it is $10\times 10\times 10\times 10$.
We diagonalize the systems with periodic boundary conditions,
obtain the eigenfunctions at the center of the energy spectrum,
input  squared modulus of the eigenfunctions to the CNN, 
and let CNN calculate the probabilities for
the inputs being delocalized.
The results are shown in Fig.~\ref{fig:phaseDiagrams} (a) (3D) and (c) (4D)
as a heat map.

We next analyze the time series of $p^2(t)$, Fig.~\ref{fig:p2t}.
We first note that
\begin{equation}
\label{eq:p2}
p^2(t)\sim\left\{\begin{array}{ll}
D t &\mathrm{delocalized}\,,\\
\xi^2 &\mathrm{localized}\,,\\
t^{2/d} &\mathrm{critical}\,,
\end{array}
\right.
\end{equation}
with $D$ the diffusion constant, $\xi$ the localization length, and $d$ the dimension.
At the critical point, $p^2(t)\propto t^{2/d}$, so for 3D $p^2(t)\propto t^{2/3}$ and 
for 4D $p^2(t)\propto t^{1/2}$\cite{Ohtsuki97}.
Note that we discuss here the localization/delocalization in momentum space.

We find the critical strength $(K,\epsilon)$ by detecting the behaviors $p^2(t)\propto t^{2/d}$,
and use this information for supervised training of LSTM.
The values of $p^2(t)$, however, strongly depend on $K$ and $\epsilon$, and the neural network
tends to learn only maxima and minima of $p^2(t)$.
We therefore preprocessed the data
by normalizing them, i.e., normalize $p^2(t)$ to $x^{(t)}$, whose
mean and standard deviation are 0 and 1, respectively.

We first determine the critical point along a straight line in
$\epsilon$-$K$ plane by finding a point that shows $p^2(t)\propto t^{2/d}$
(see white crosses in Fig.~\ref{fig:phaseDiagrams}).
Once the critical point is determined, we prepare time series $p^2(t)$ for
localized and delocalized phases by varying $(K,\epsilon)$ along
a straight line indicated by green arrows in Fig.~\ref{fig:phaseDiagrams}(b),(d).
We then  normalize $p^2(t)$ to $x^{(t)}$ and use them for training
bidirectional LSTM.
Once the LSTM is trained, we vary parameters in $\epsilon$-$K$ plane and
calculate $x^{(t)}$, and feed $x^{(t)}$ to LSTM, which outputs the
probability that the input time series $x^{(t)}$ belongs to the delocalized phase.
The results are shown in Fig.~\ref{fig:phaseDiagrams}(b) for 3D and Fig.~\ref{fig:phaseDiagrams}(d) for 4D,
which nicely distinguish localized and delocalized phases.

Now we compare the phase diagrams (heat maps) for 3D and 4D systems.
In the case of 3D, both CNN and LSTM give reasonably sharp phase boundaries
[see Fig.~\ref{fig:phaseDiagrams}(a), (b)].
On the other hand, in the case of 4D, the phase boundary becomes  noisy
if we use 4D CNN [Fig.~\ref{fig:phaseDiagrams}(c)].
This is because in the case of 4D, only small system can be diagonalized, and
CNN fail to learn the detailed features of localized and delocalized states.
In the case of LSTM [Fig.~\ref{fig:phaseDiagrams}(d)], the phase boundary is less noisy,
since we do not need to diagonalize the system, and we can follow as long time series as in
3D case.

\section{Summary}
To summarize, we have analyzed the quantum kicked rotor with time modulated kick strength.
The systems are analyzed in two ways.  One is to map the Hamiltonian
to static higher dimensional
tight binding models and study the eigenfunctions via the deep convolutional neural network.
The other is to analyze the time series of the original time dependent one dimensional systems
via bidirectional long short-term memory
network.  We have demonstrated that the latter approach gives less noisy phase boundary
between the localized and delocalized phases. 
The latter approach works especially well for analyzing the Anderson transitions in
higher dimensions.

\medskip
\noindent
{\bf Acknowledgement}
This work was supported by JSPS KAKENHI Grant Nos. 16H06345, and 19H00658.
We thank Dr. Matthias Stosiek for critical reading of the manuscript.

\bibliography{manoOhtsuki21.bib}

\end{document}